\pdfoutput=1
\documentclass{JINST}
\usepackage{graphicx}

\title{A new application of emulsions to measure the gravitational force on antihydrogen}

\author{C. Amsler$^a$, A. Ariga$^a$, T. Ariga$^a$, S. Braccini$^a$, C. Canali$^b$, A. Ereditato$^a$, J. Kawada$^a$, M. Kimura$^a$, I. Kreslo$^a$, C. Pistillo$^a$, P. Scampoli$^{a,c}$\thanks{Corresponding
author.}~, J.W. Storey$^a$\\
\llap{$^a$}Albert Einstein Center for Fundamental Physics,\\
Laboratory for High Energy Physics, University of Bern,\\
Sidlerstrasse 5, CH-3012 Bern, Switzerland\\
\llap{$^b$}Physics Institute, University of Zurich,\\
Winterthurerstrasse 190, CH-8057 Zurich, Switzerland\\
\llap{$^c$}Department of Physical Sciences, University of Napoli Federico II,\\
  Complesso Universitario di Monte S. Angelo, I-80126, Napoli, Italy\\
  E-mail: \email{paola.scampoli@lhep.unibe.ch}}

\abstract{We propose to build and operate a detector based on the emulsion film technology for the measurement of the gravitational acceleration on antimatter, to be performed by the AEgIS experiment (AD6) at CERN. The goal of AEgIS is  to test  the weak equivalence principle with a precision of 1\% on the gravitational acceleration $g$ by measuring the vertical position of the annihilation vertex of antihydrogen atoms after their free fall in a horizontal vacuum pipe. 
With the emulsion technology developed at the University of Bern we propose to improve the  performance of  AEgIS by exploiting the superior position resolution of emulsion films over other particle detectors. 
The idea is to use a new type of emulsion films, especially developed for applications in vacuum, to yield a spatial resolution of the order of one micron in the measurement of the sag of the antihydrogen atoms in the gravitational field. This is an order of magnitude better than what was planned in the original AEgIS proposal.}

\keywords{Particle tracking detectors; emulsion detector; antihydrogen; gravitational acceleration; AEgIS}

\begin{document}

\section{Introduction}

The  AEgIS experiment at CERN~\cite{aegisprop} was designed to directly measure for the first time the Earth's local gravitational acceleration for antimatter, namely antihydrogen ($\bar H$) atoms, with the aim of testing the weak equivalence principle (WEP). Electrically neutral $\bar H$ atoms are ideal to probe the WEP since measurements conducted so far with the available charged antiparticles are affected by the overwhelming electromagnetic background, which prevents the determination of the gravitational acceleration. AEgIS is expected to start taking data in 2014-2015 at the Antiproton Decelerator (AD) at CERN, the only laboratory in the world  providing low energy antiproton beams suitable for the production of $\bar H$ atoms.

	The WEP states that the trajectory of a point mass experiencing the gravitational field is completely defined by its initial position and velocity, regardless of its composition. This has been investigated for matter with a relative precision down to $10^{-13}$ by E\"otv\"os-type experiments with various probes, but never with antimatter. The measurement proposed by the AEgIS collaboration~\cite{aegisprop} will be performed by measuring the vertical position of horizontally launched $\bar H$ atoms after their free fall in a vacuum tube. The initial goal of the experiment is to reach a relative precision $\Delta g/g$ of 1\% on the measurement of the gravitational acceleration, which requires a position resolution of about $10~\mu$m for a position sensitive detector that has to measure the annihilation point at the end of the $\bar H$ atom trajectory.
	
	We investigate here the use of an emulsion-based position detector as an alternative to the originally foreseen silicon micro-strip detector. Emulsion films, with a position resolution at the sub-micrometric level~\cite{handbook}, can  be used as tracker devices to detect the $\bar H$ annihilation vertex, or as a target, by directly observing the annihilation vertices in the emulsions. With emulsion films the resolution will be appreciably better than  $10~\mu$m. Nevertheless, an intense R\&D programme is required on emulsion films, since the detectors will operate in unprecedented conditions. In particular, they will be used for the first time in vacuum.
	 In addition, the data acquisition and analysis software for the emulsion detectors we are presently employing have been developed for the OPERA neutrino oscillation experiment~\cite{opera}. In OPERA particles produced by high energy neutrinos which hit the emulsion detectors are mainly emitted in the forward direction. Scanning procedures and reconstruction software were therefore designed to maximize the performance at small angles. On the contrary, the angular distribution of the annihilation products from stopping $\bar H$ atoms is  isotropic. Furthermore, the event reconstruction algorithms in OPERA mostly relies on Minimum Ionizing Particle (MIP) detection, while for AEgIS the detection of nuclear fragments and highly ionizing particles plays an important role. For these reasons an upgrade of the scanning system is necessary and two R\&D projects, one aiming at the adaptation of the current scanning procedure and the other at the design of completely new procedures and algorithms for track reconstruction, are being conducted by our group in parallel with the necessary studies related to the emulsion operations in vacuum.

\section{AEgIS and its experimental requirements}
\begin{figure}[htbp]
        \centering
        \includegraphics[height=30cm, width=11cm, angle=0,keepaspectratio]{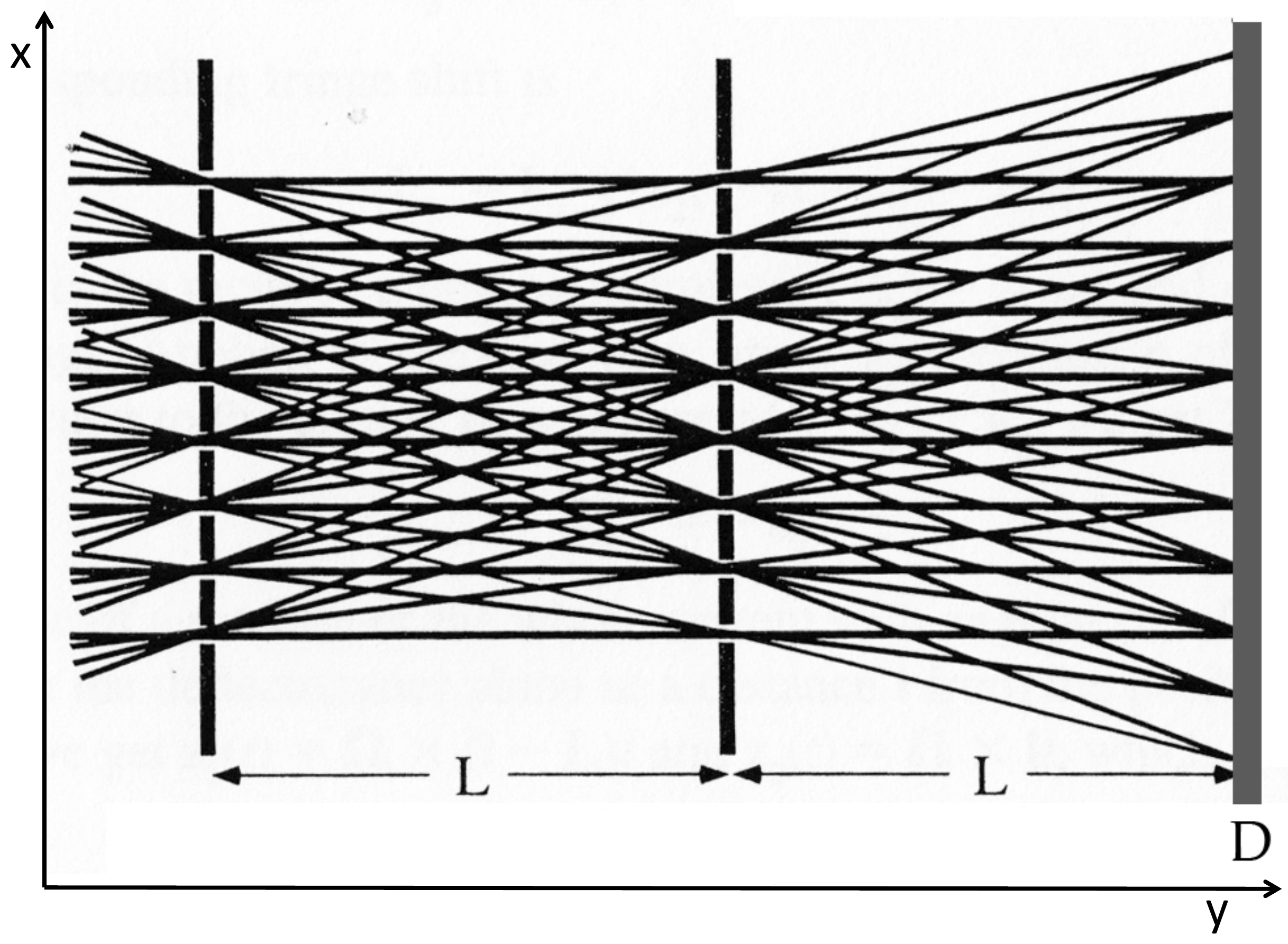}
        \caption{\emph {Moir\'e deflectometer \cite{batelaan} and a position sensitive detector (D).}}
        \label{moire}
\end{figure}

The $\bar H$ atoms are produced through the interaction of cold antiprotons with positronium atoms excited to a Rydberg state (Ps*), resulting in the capture of the bound positron by the antiproton and in the release of an electron, according to the charge exchange reaction $Ps^*+\bar p \rightarrow \bar H^*+e^-$~\cite{charlton}. $\bar H$ atoms are produced with an average kinetic energy corresponding to a temperature of 100 mK. The $\bar H$ atoms in the Rydberg states can be Stark-accelerated along the beam axis by an inhomogeneous pulsed electric field to a horizontal velocity of a few 100 m/s~\cite{vliegen}, and are then allowed to fall freely over a distance of typically 1 m, traversing a Moir\'e deflectometer \cite{batelaan}. Such a device is already used for gravity measurements on argon beams~\cite{oberthaler}. For AEgIS the deflectometer will consist of  two identical gratings placed at a distance $L$  of 40 cm (figure~\ref{moire}). The intensity pattern would be recorded along the vertical direction $x$ on a position sensitive detector at the same distance $L$ from the second grating. The intensity pattern shows the same periodicity as the gratings. The downward (or upward) displacement $\Delta x$ of the Moir\'e  intensity pattern due to gravity can then be measured with respect to the case with no gravity. Since the pulsed $\bar H$ beam is not monochromatic, $\Delta x$ depends on the time of flight of the $\bar H$ atoms through the relation $\Delta x$ = $g T^2$, where $T$ is the time of flight between the two gratings. Thus the position detector has to be coupled to a time of flight device detecting the annihilation products. 
The time of flight can be computed from the switch off time of the electric field for Stark acceleration and the $\bar{H}$ annihilation time. For a detailed description of the AEgIS experiment 
see ref.~\cite{aegisprop}. 

A crucial issue for the proposed measurement is the number of $\bar H$ atoms traversing the deflectometer and reaching the final detector. The Moir\'e intensity pattern is related to the geometry of the Moir\'e system (grating pitch and grating width). The vertical intensity distribution (I) on the position detector depends on the time of flight between the gratings and on the contrast $C =(I_{max} - I_{min})/(I_{max} + I_{min})$, directly connected to the detector resolution~\cite{oberthaler}.
The position sensitive detector would cover an area of about $20 \times 20$ cm$^2$ to match the geometry of the Moir\'e deflectometer and measure the vertical coordinate of the annihilation vertices with a precision of  $\sigma \simeq10~\mu $m. The baseline option considered by AEgIS is a silicon micro-strip detector, extensively discussed in~\cite{bonomi}. The present paper describes developments performed at the Laboratory for High Energy Physics (LHEP) of the University of Bern to improve on the position resolution by a factor of ten.
     
\section{Emulsion detector developments}

\begin{figure}[htbp]
        \centering 
        \includegraphics[width=0.6\textwidth, height=0.5\textwidth]{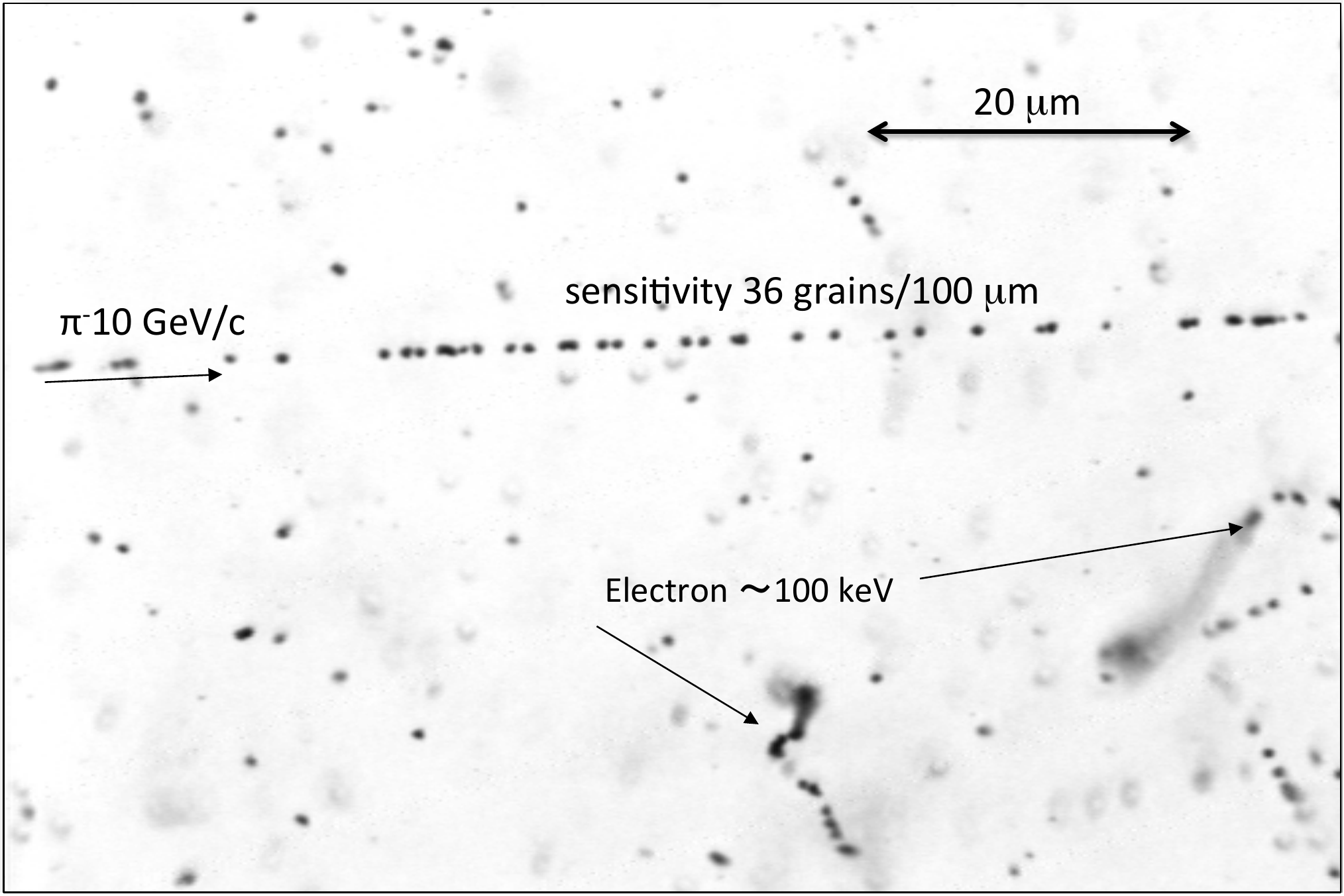}
        \caption{\emph {A MIP track from a 10 GeV/c pion in an emulsion.}}
       \label{operafilm}
\end{figure}

\begin{figure}[htbp]
      \centering
     \includegraphics[width=0.6\textwidth, height=0.5\textwidth]{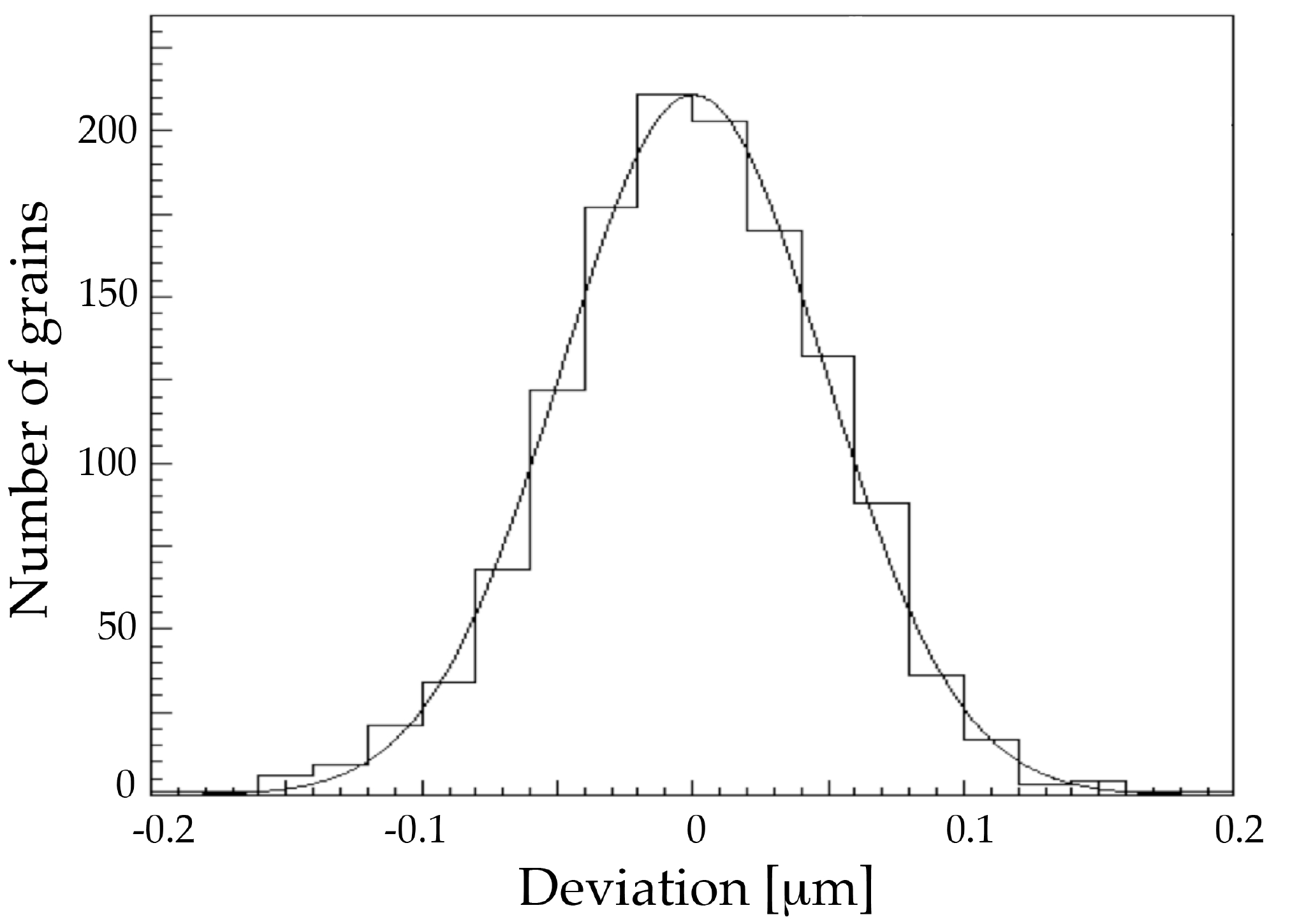} 
    \caption{\emph {Deviation of grains from a linear-fit line for horizontal tracks.}} 
    \label{resolution}
\end{figure}

       Nuclear emulsions are among the first particle detectors that led physicists to important discoveries, thanks to their unbeatable sub-micrometric position resolution. Today's nuclear emulsions are based on the modern, industrial photographic film technology developed in the last years~\cite{handbook}. The basic detector unit is a silver bromide crystal with a typical size of 0.2 $\mu$m, uniformly distributed in an emulsion layer, and characterized by an energy band gap of 2.6 eV. The energy loss of a charged particle causes the excitation of the carriers, which are successively trapped in intermediate states at the lattice defects on the crystal surface, hence forming latent images. A developing treatment can produce a filament of pure silver visible under an optical microscope as a grain of $\sim$0.6 $\mu$m diameter. A full three-dimensional reconstruction of particle tracks is possible even for very thin emulsion layers ($\sim$50 $\mu$m), by evaluating the silver grain density along the particle track. In the past, the main limiting factor in the use of these detectors was the time consuming manual scanning procedure. The recent, impressive development of fast automated scanning systems~\cite{aoki}~\cite{morishima} has fostered a renewed interest in this technique, particularly in the field of neutrino physics.
       A big step forward has been done in Japan on the development of photographic emulsion detectors, in particular by the Nagoya University, in collaboration with the Fuji Company for the needs of the OPERA neutrino oscillation experiment~\cite{nakamura}. A standard industrial emulsion film is made of two gelatine layers with a thickness of 44 $\mu$m coated on both sides of a plastic base of 205 $\mu$m thickness. A track produced by a charged particle is detected as a sequence of silver grains, as shown in figure~\ref{operafilm}, where about 35 silver grains per 100 $\mu$m are created by a MIP.  The intrinsic spatial resolution is about 0.050 $\mu$m, as shown in figure~\ref{resolution}.
  
	An emulsion film of 100 cm$^2$ surface consists of about $10^{14}$ channels of silver bromide crystal detectors, each with a detection efficiency of about 16\% for a MIP. The read-out of this enormous number of channels is performed by optical microscopes driven by state-of-the-art fast electronics. 
The present use of this technique has been made possible by the impressive development of automated scanning systems started in 1982 at the Nagoya University by K. Niwa and collaborators, and successfully used in neutrino physics experiments such as CHORUS~\cite{chorus} and DONUT~\cite{donut}. This system evolved its scanning ability from 0.2 microscope views per second in 1982 to 30 views per second in 2001 (one standard view is 150 $\times$ 120 $\mu$m$^2$). 
An independent scanning system was developed in Europe, the so-called European Scanning System (ESS), able to reach a frame rate of 400 frames per second and to scan up to 20 cm$^2$/hour of emulsion surface with real-time track reconstruction~\cite{ess1}\cite{ess2}. 

\begin{figure}[htbp]
        \centering
        \includegraphics[height=30cm, width=12cm, angle=0,keepaspectratio]{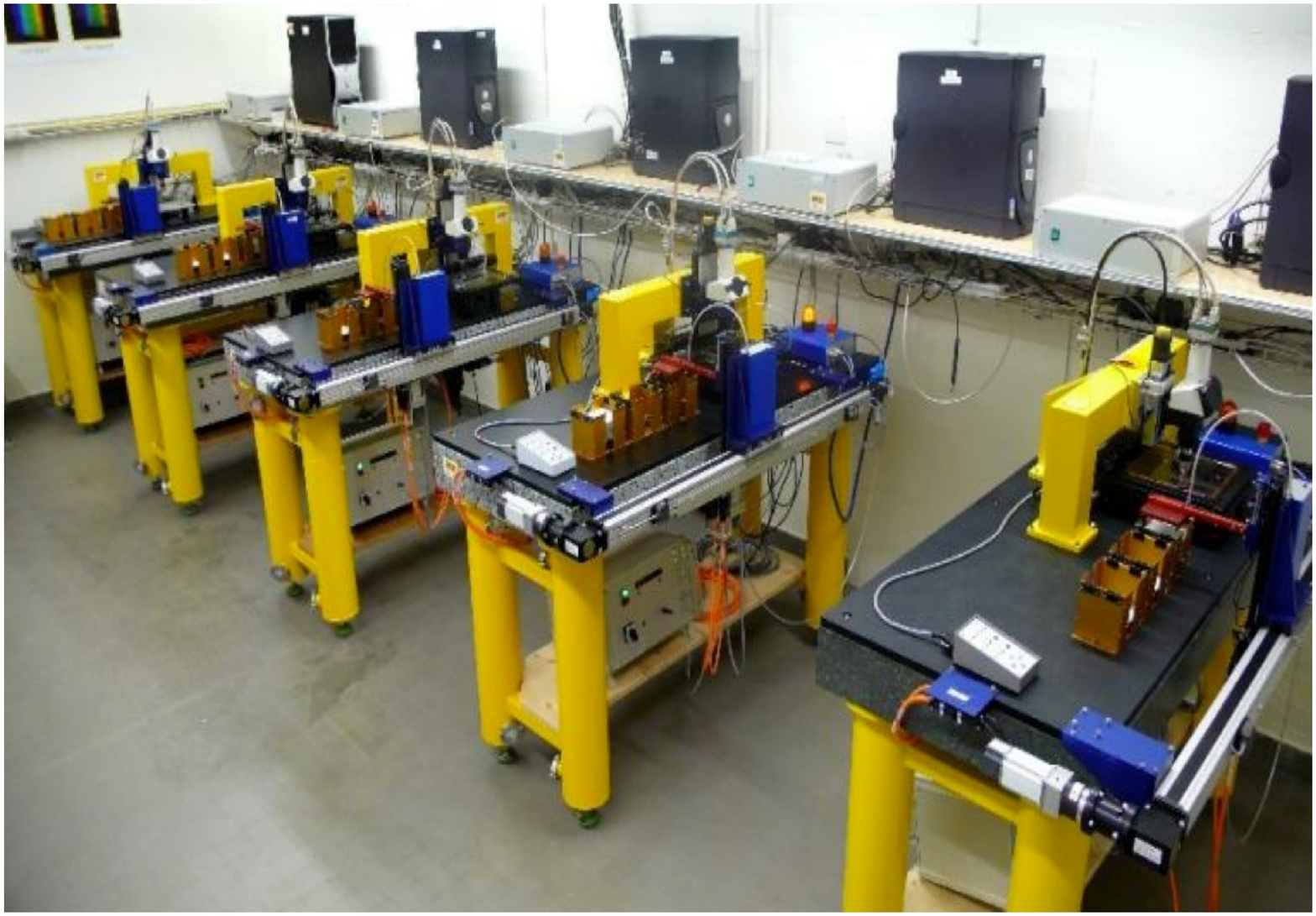}
        \caption{\emph {Emulsion scanning facility at the LHEP in Bern.}}
        \label{microscopes}
\end{figure}

	The LHEP played a key role in the development of the ESS, with the implementation of the dry objective scanning technique~\cite{kreslo} and  the realization of robots for the automatic emulsion film handling~\cite{borer}.  LHEP presently hosts the largest scanning laboratory in Europe with its six microscopes for the OPERA event analysis, the second in the world, after that of Nagoya (figure~\ref{microscopes}). One microscope is completely devoted to the R\&D activity on emulsions.
	
Furthermore, a dedicated emulsion laboratory is also operational at LHEP. It is located 30 m underground since a low cosmic-ray track background is needed for the emulsion film production and development. Temperature and humidity are kept stable at 18$^\circ$C -- 20$^\circ$C, and 50\% -- 60\%, respectively, as required for the handling and the development of emulsions. Custom made films can then be produced to cope with specific experimental requirements. A high precision stage to pour the sensitive film on the supporting plastic (or glass) base and equipped with a vacuum system permits the production of custom emulsion films with sizes up to 20 $\times$ 30 cm$^2$. The LHEP emulsion development capability presently reaches about 1 m$^2$ of films per day.

\section{An emulsion detector for the AEgIS experiment}

In the  AEgIS proposal \cite{aegisprop} a silicon micro-strip detector with a 20 $\times$ 20 cm$^2$ sensitive area (20 $\mu$m strip pitch and  300 $\mu$m thickness) was considered as detector to measure the vertical coordinate of the $\bar H$ annihilation point with an r.m.s. of 10 $\mu$m. Additional planes of silicon strip detectors behind the first plane were also foreseen to reconstruct the trajectories of pions emerging from the $\bar H$ annihilations.
Indeed, the reconstruction of the annihilation vertex is essential to reduce the background from annihilations elsewhere in the apparatus, e.g. in the gratings.

We propose an alternative design that features a nuclear emulsion tracking detector which improves by one order of magnitude (from 10 $\mu$m to 1 $\mu$m) the expected position resolution, thus obtaining a significant reduction of the uncertainty on the measurement of $g$ and/or a strong reduction of the running time.

\begin{figure}[htbp]
        \centering
        \includegraphics[height=30cm, width=9cm, angle=0,keepaspectratio]{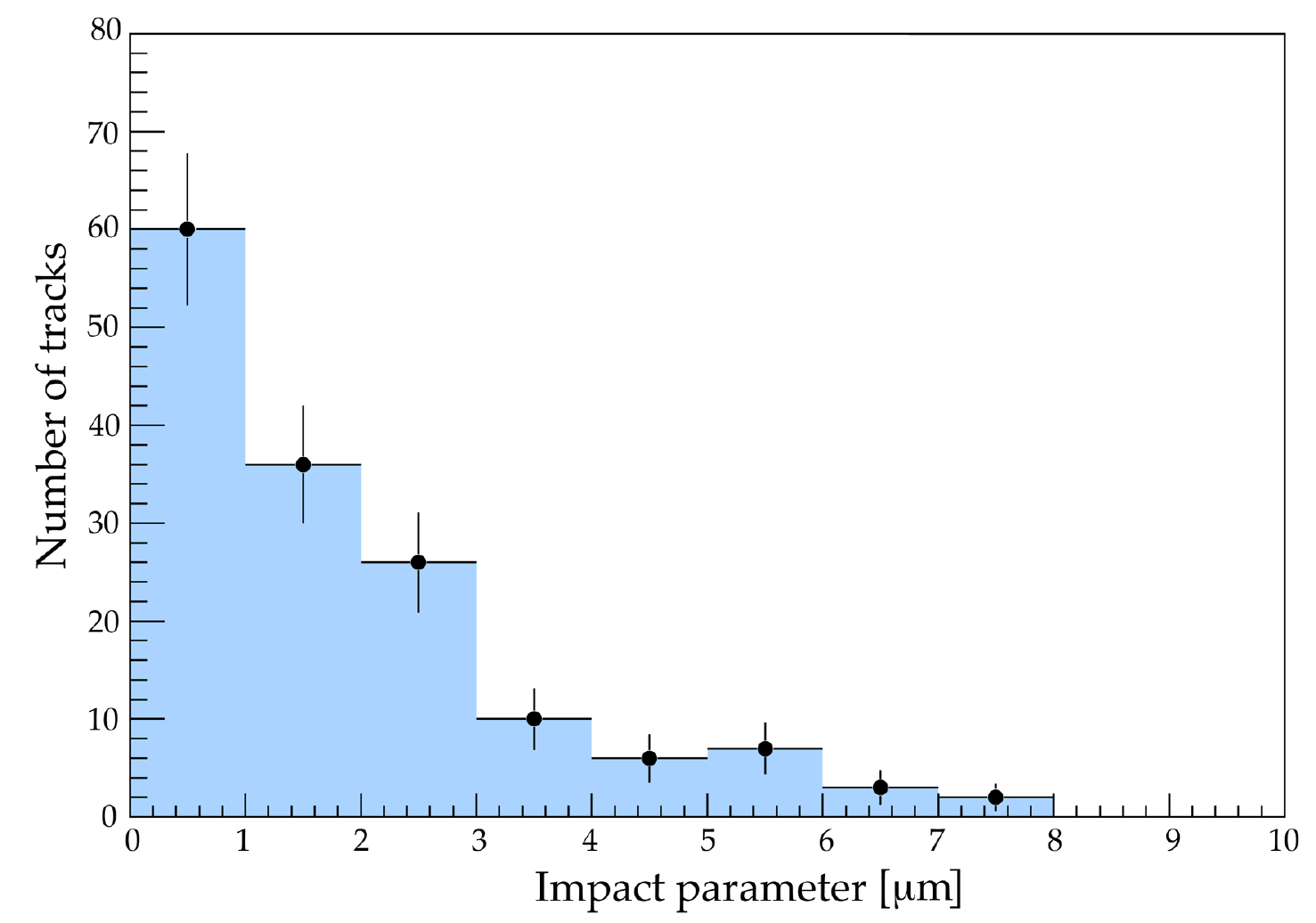}
        \caption{\emph {Impact parameter distribution (IP) for tracks from $\pi$ interactions in 1 mm lead (8 GeV/c $\pi$ beam at CERN).}}
        \label{ip}
\end{figure}

	The emulsion film tracker will operate at room temperature and will be located in the low vacuum region ($\sim$$10^{-6}$ mbar) of the AEgIS apparatus, immediately behind a thin window needed to separate it from the ultra-high vacuum region ($10^{-11}$ mbar) in the beam pipe hosting the deflectometer. The $\bar{H}$ atoms annihilate on the surface of the foil which has to have a minimal thickness in order to minimize multiple scattering. 
The position and angular resolution provided by the emulsions allow to detect particles produced in the annihilation vertex (mainly pions, protons and nuclear fragments) and to reconstruct vertices with high precision, that is with a track impact parameter (defined as the minimum distance between the track and the reconstructed vertex) well below 10 $\mu$m. As an example, we show in figure~\ref{ip} the measured impact parameter distribution for tracks emerging from pion interactions  with a 1 mm thick lead plate. This emulsion exposure was performed in 2004 at CERN to study hadron re-interactions. A detector made of emulsions interleaved in 1 mm thick lead plates was exposed to a low intensity 8 GeV/c pion beam. It should be noted, however, that for such relatively thick plates multiple scattering is much more severe than for the thin window needed in AEgIS.

The emulsion tracker has to be positioned as close as possible to the $\bar H$ annihilation surface (ideally closer than 0.1 mm to the window) to minimize the error in extrapolating the  track projection to the production point.

\begin{figure}[htbp]
        \centering
        \includegraphics[height=30cm, width=15cm, angle=0,keepaspectratio]{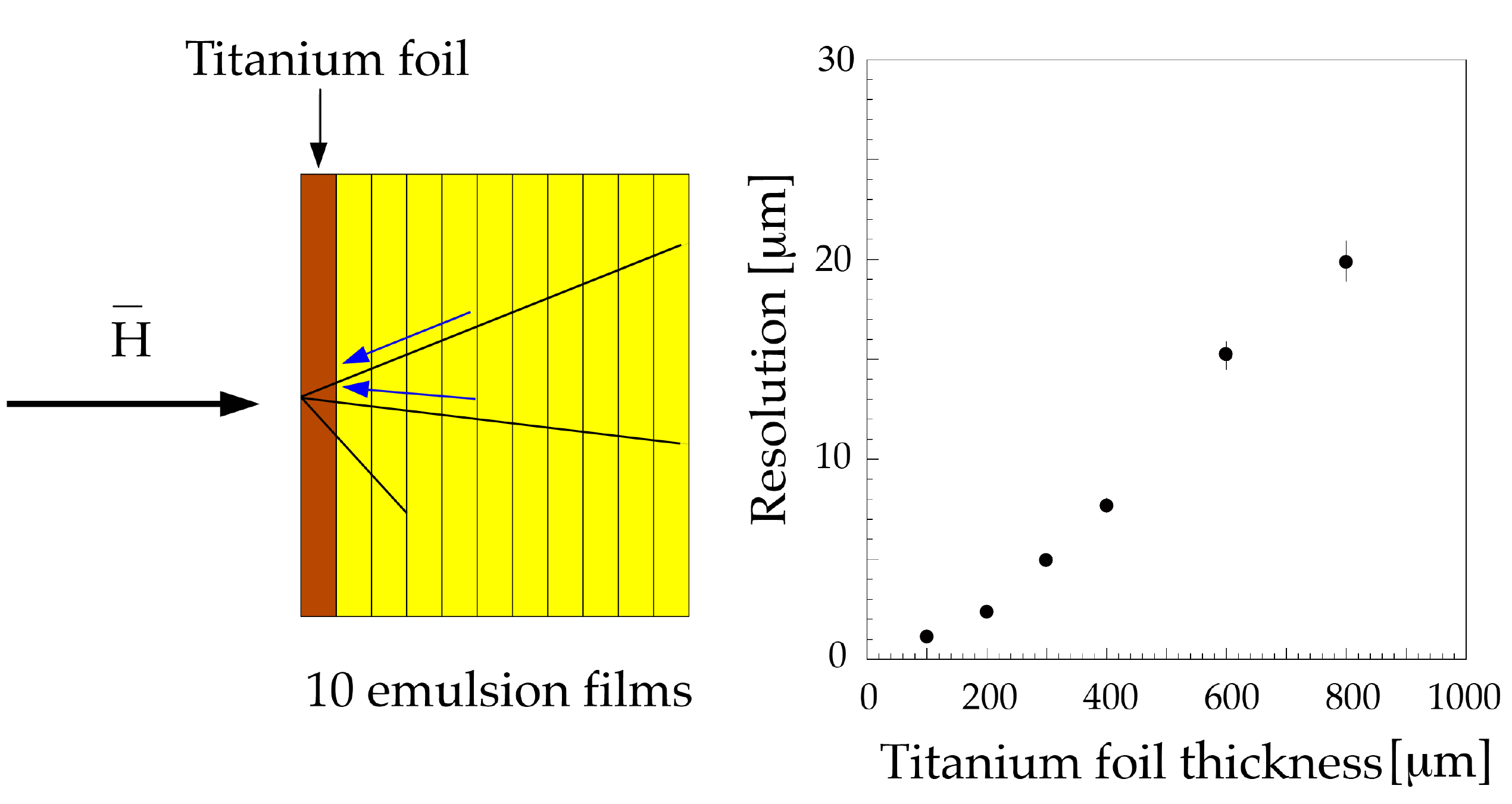}
        \caption{\emph {Sketch of the simulated emulsion stack (left). R.m.s annihilation vertex resolution vs. titanium window thickness (right).}}
        \label{mcm}
\end{figure}

A Monte-Carlo simulation (GEANT4 package supplied with the QGSP\_BERT hadronization model~\cite{geant}) was performed to estimate the accuracy of the reconstructed interaction point  as a function of the thickness of a metal foil, such as titanium, placed in front of the emulsion films. 
A possible setup with a stack of ten emulsion films to improve track reconstruction is shown in figure~\ref{mcm}.
Emitted tracks from $\bar p$ annihilation mainly consist of charged pions, protons, and nuclear fragments. Since the average momentum of pions and protons is around 0.3 GeV/c, one can identify particles from the level of  track darkness in the emulsion given by their different $dE/dx$~\cite{toshito}. On the average 1.2 pions and 1.7 protons per annihilation pass through a 100 $\mu$m thick titanium foil. The annihilation vertex is then reconstructed from the tracks in the emulsions. Annihilation points are determined by extrapolating tracks to the titanium foil.
The results are shown in figure~\ref{mcm} (right). With a 100 $\mu$m thick titanium foil, one can expect a position resolution of the order of 1 $\mu$m. Preliminary tests with low energy antiprotons stopping in the foil have been performed in summer 2012 \cite{pbarresults}.

The emulsions will have probably to be replaced during beam exposure. The maximum accumulated flux, still manageable in the offline analysis, is around 10$^5$/cm$^2$. Preliminary estimates based on the background \emph{in situ} lead to a lifetime of the emulsion films of the order of a few days before they need to be replaced. \emph{Ad-hoc} devices allowing their insertion and removal will be designed. Such operations are relatively affordable for a passive detector as emulsions that do not need any power supply nor readout electronics. 

A time of flight hodoscope placed behind the emulsion tracker will also be needed to measure the $\bar H$ atoms arrival time and to determine the approximate position of particles emerging from $\bar H$ annihilations in the emulsions. The time of flight and the approximate track locations will be stored for the offline analysis of the emulsion films. A scintillator fiber tracker or a silicon micro-strip detector are  being considered. The time of flight hodoscope should have adequate position and angular resolution to avoid ambiguities when matching emulsion data. 
A silicon strip detector of suitable thickness can also be considered as an active separation window. With this configuration, the employment of two complementary technologies would allow for a continuous cross-check of data quality and an overall reduction of systematic uncertainties. 

We have performed a Monte-Carlo simulation of the expected performance of the emulsion detector assuming a realistic  setup for the AEgIS experiment~\cite{aegisprop}, in particular a distance $L$ of 40 cm, a grating pitch of 80 $\mu$m and slit sizes of 24 $\mu$m, corresponding to an open fraction of 30\%. 
Assuming  $g = 9.8$ m/s$^2$, and fitting the $\bar H$ atoms vertical displacements as a function of the time of flight between the two gratings ($\Delta x = g T^2$) a 1\% (1.5\%) precision for $\Delta g/g$ was found for 1 $\mu$m (10 $\mu$m) resolution for  7000 detected $\bar{H}$ atoms (1000 $\times$ 7 time of flight bins), in good agreement with the simulation reported in the AEgIS proposal. It is worth noting that with 1 $\mu$m position resolution we obtained the same precision on $\Delta g/g$ as the one exhibited by an ideal detector with infinite precision. 

\begin{figure}[htbp]
        \centering
        \includegraphics[height=20cm, width=11cm, angle=0,keepaspectratio]{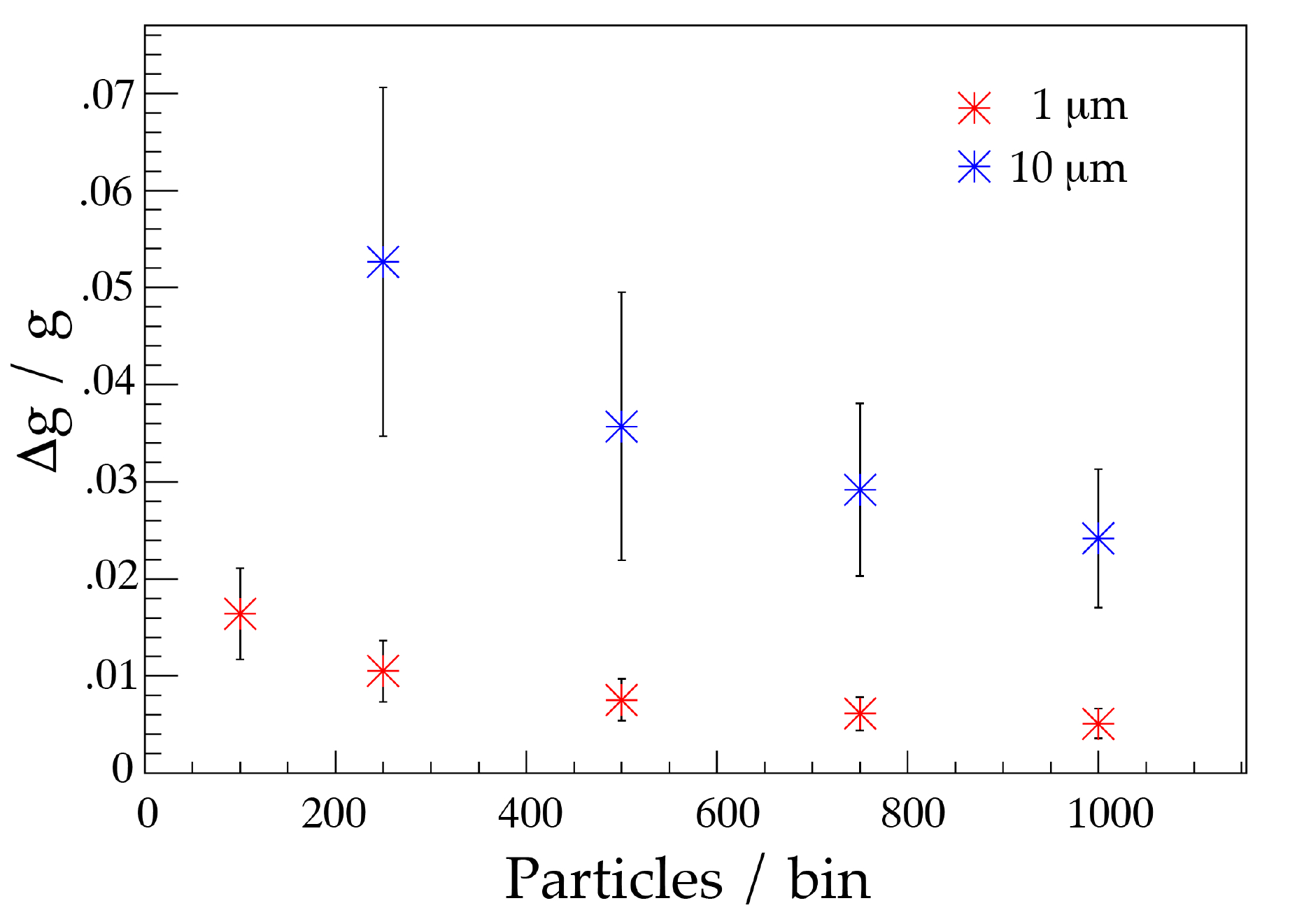}
        \caption{\emph {$\Delta g/g$ vs. number of detected particles for a position sensitive detector resolution of 1 $\mu$m (red) and 10 $\mu$m (blue) for 40 $\mu$m pitch and 30\% open fraction  Moir\'e deflectometer. The particles were grouped into 7 time of flight bins.}}
        \label{c4}
\end{figure}

However, a deflectometer with a pitch of 80 $\mu$m and an open  fraction of 30\% optimizes the performances of a 10 $\mu$m resolution detector but  does not exploit the potentialities of a detector featuring one order of magnitude better resolution. With a different deflectometer geometry, for example with a  pitch of 40 $\mu$m and  30\% open fraction, the performances of a position resolution of 10 $\mu$m detector worsen, while a precision of 0.5\% can be achieved with 1 $\mu$m position resolution. These results are shown in figure~\ref{c4} where $\Delta g/g$ vs. the number of detected particle is reported for a 40 $\mu$m pitch deflectometer. The $\Delta g/g$ value for 100 detected particles per bin is not quoted for the 10 $\mu$m position resolution since the adopted analysis method fails when fitting $g(\Delta x, T)$. For 1 $\mu$m position resolution, with 250 particles per bin, the 1\% precision on $\Delta g/g$ is achieved with less that 2000 detected annihilations. This corresponds to a factor of 4 reduction in beam time, so that the requested precision is guaranteed should the number of produced $\bar H$ atoms be smaller than expected. 

\section{R\&D on emulsions for high precision applications in vacuum}
\begin{figure}[htbp]
        \centering
 	  \includegraphics[height=9cm]{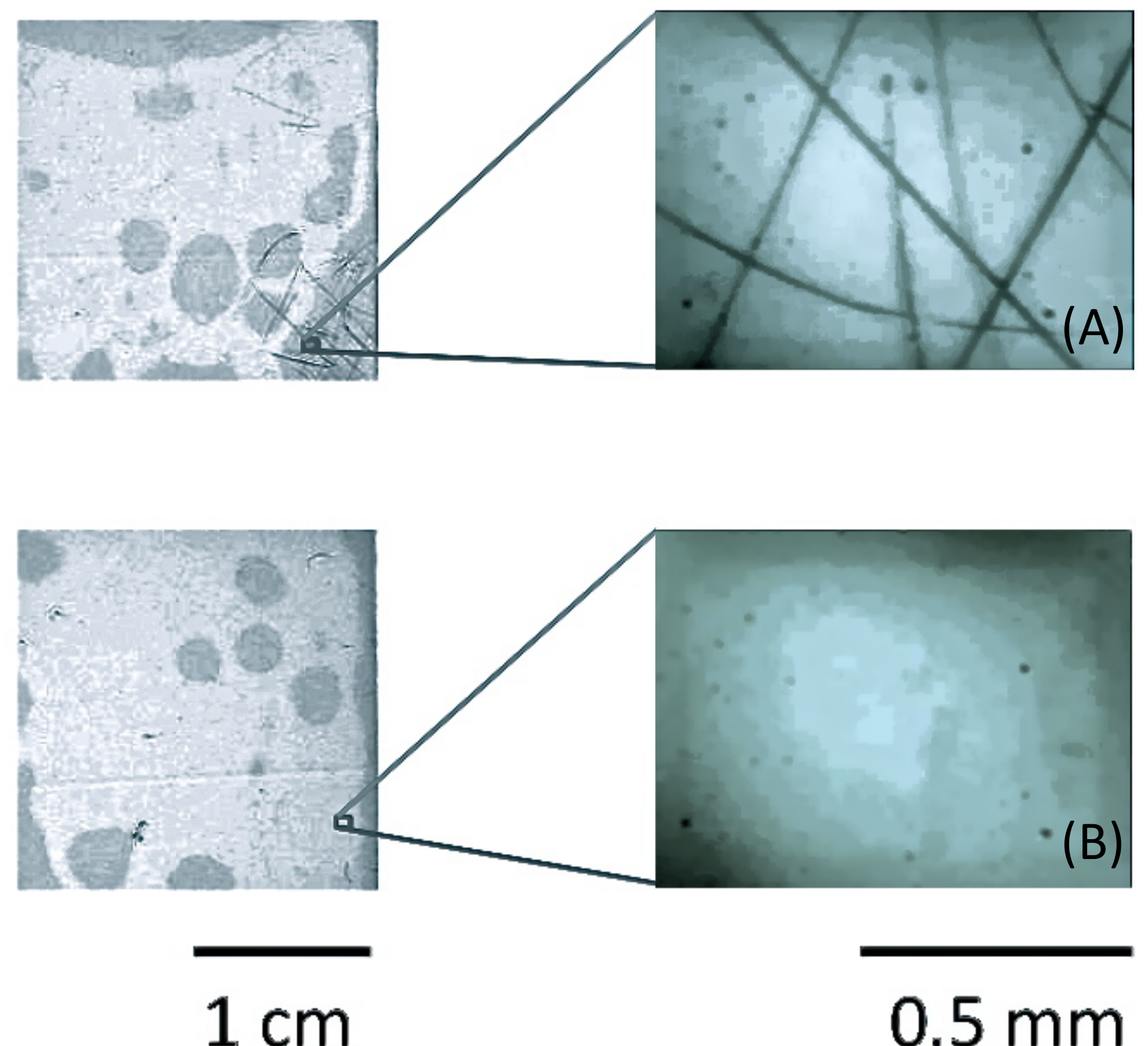}     
        \caption{\emph {Emulsion films after 3.5 days in the vacuum chamber without glycerine treatment (A) and with the glycerine treatment (B): in the first case cracks in the emulsion are clearly visible while they disappear after treatment  with 17\% glycerine solution. A view of the film surfaces observed with a microscope is shown on the right.}}
        \label{cracks}
\end{figure}

Tests were performed to study the behavior of emulsion detectors in vacuum, in particular as far as mechanical stress, background (density of thermal induced grains, called "fog") and sensitivity (Grain Density (GD) along the track trajectory) are concerned. We studied the properties of OPERA type films in a vacuum chamber at a pressure of 10$^{-6}$ mbar. The plastic base and two emulsion layers of the films have a considerable water content. The water loss under vacuum conditions, in particular in the emulsion gelatine, can produce cracks in the emulsion layers, hence altering the relative distance of tracks that have to be measured at the $\mu$m level. Cracks occurrence in emulsions was evaluated for films treated with different glycerine solutions, since glycerine can prevent the elasticity loss of the detector. The treatment consists in dipping the films in solutions with different glycerine concentration for a time ranging from 20 to 60 minutes. After 3.5 days in the vacuum chamber the treatment with glycerine can effectively prevent cracks, as shown in figure~\ref{cracks}.

\begin{figure}[htbp]
       \centering 
          \includegraphics[height=9cm, width=12cm, angle=0,
         keepaspectratio]{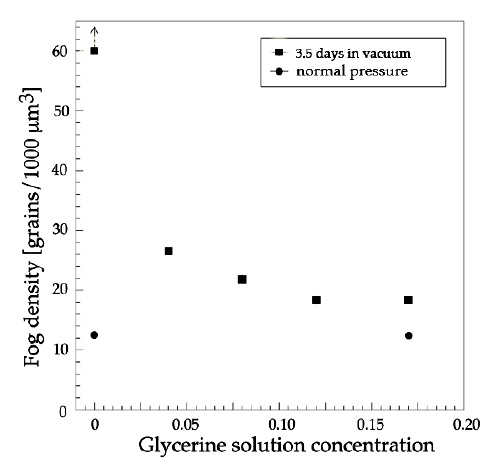}
        \caption{\emph {Fog density (FD) vs. glycerine solutions for films kept in vacuum chamber for 3.5 days (squares) and for films at room temperature and pressure (dots).}}
       \label{fog}
\end{figure}

	The Fog Density (FD) ranges from 10 to 13 grains per 1000 $\mu$m$^3$ for OPERA type detectors. It was measured for the developed detectors after a 60 minute treatment with glycerine solutions, both in normal environment and after having been kept in vacuum. For this measurement films were dipped in different glycerine solutions (while a reference film was dipped in de-ionized water), dried for 2 days at a relative humidity of 60\% before being placed in the vacuum chamber for 3.5 days at 10$^{-6}$ mbar. After this procedure, they were developed following the protocol of the OPERA experiment.
In figure~\ref{fog} the measured FD is reported as a function of glycerine content (4\%, 8\%, 12\% and 17\%) in vacuum, and at  0\% and 17\%  glycerine content at room temperature and pressure. In the latter case, background results are almost unaffected by glycerine, while the films with vacuum treatments show increased FD values. In particular, the films that had not been treated with glycerine showed a considerable background increase in vacuum, with an FD value larger than 60 grains/1000 $\mu$m$^3$, but difficult to quote due to the considerable blackening of the films. On the other hand, the background was much reduced for treated films. An optimal concentration for the glycerine solution cannot be defined yet. Our measurements suggest that a film treatment with a $\sim$10\% glycerine solution can maintain the background at levels we can cope with. In any case, developments of new gels will be certainly  beneficial to improve the FD. 

\begin{figure}[htbp]
       \centering 
          \includegraphics[height=9cm, width=14cm, angle=0,
         keepaspectratio]{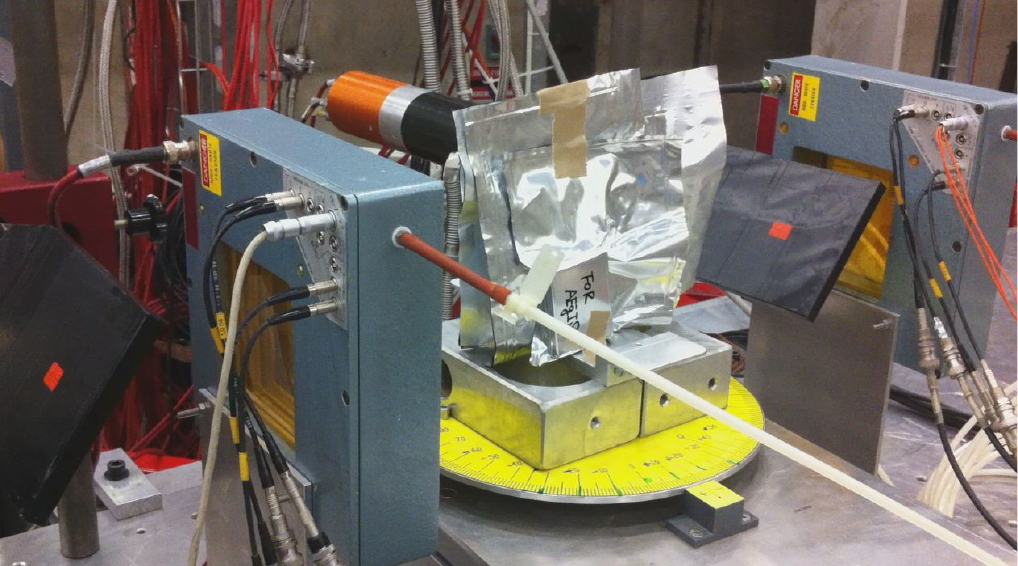}
        \caption{\emph {Exposure setup for the crystal sensitivity measurement (see text).}}
       \label{2012_test}
\end{figure}

	The sensitivity of silver bromide crystals in  glycerine-treated OPERA films was checked in vacuum by using MIPs. The crystal sensitivity is defined as the number of grains observed after photo-development over the number of crystals crossed by a MIP. Films were dipped in four solutions with different glycerine concentrations for 1 hour, and put in the vacuum chamber for 1 day (at $\sim$10$^{-5}$ mbar) or 3 days (at $\sim$10$^{-6}$ mbar). The films were then exposed to a 6 GeV/c pions at the CERN PS/T10 beam-line in summer 2012. The setup is shown in figure~\ref{2012_test}. The beam profile and intensity were measured by multi-wire chambers and scintillation counters. The beam density was 10$^4$ particles/cm$^2$. The films (inside a aluminum bag) were positioned horizontally with respect to the pion beam. 
	
\begin{figure}[htbp]
       \centering 
          \includegraphics[height=9cm, width=16cm, angle=0,
         keepaspectratio]{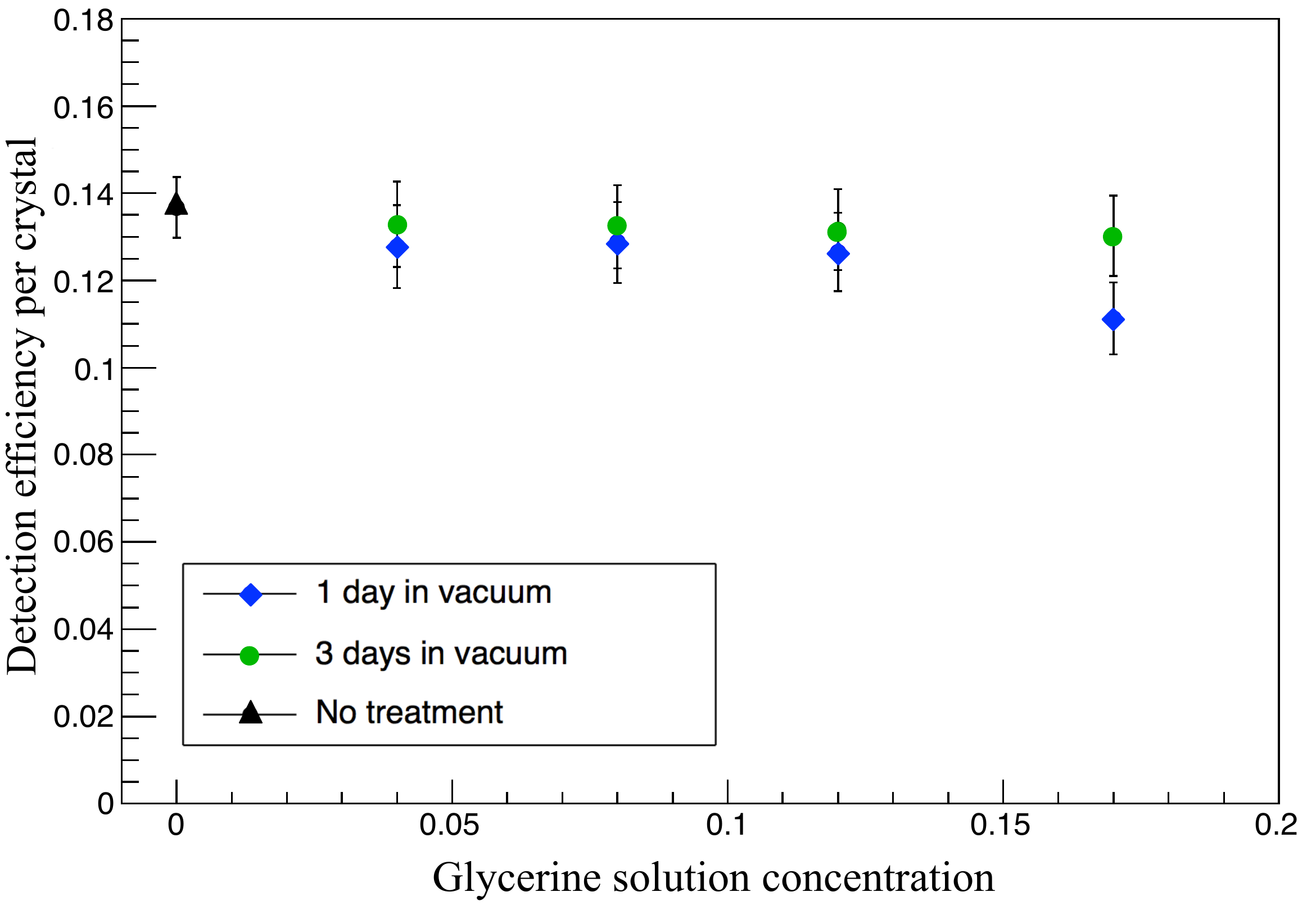}
        \caption{\emph {Crystal sensitivity vs. glycerine concentration.}}
       \label{sensitivity}
\end{figure}

	In a reference OPERA film a MIP crosses 230 crystals per 100 $\mu$m~\cite{nakamura}, and about 35 grains per 100 $\mu$m are formed after photo-development for a crystal sensitivity of 14\%. The thickness of glycerine-treated films increases with respect to the original one, and consequently the density of the crystals decreases. A correction factor, defined as  the ratio of the film thickness after the treatment over the original film thickness, is thus applied to the treated films in order to correctly estimate the crystal sensitivity. Figure~\ref{sensitivity} shows the measured crystal sensitivity as a function of glycerine content. No degradation of crystal sensitivity was found by dipping emulsions in glycerine, even after placing the films in vacuum for a period of a few days.

\begin{figure}[h!]
        \centering
          \includegraphics[height=12cm, width=12cm, angle=0,
          keepaspectratio]{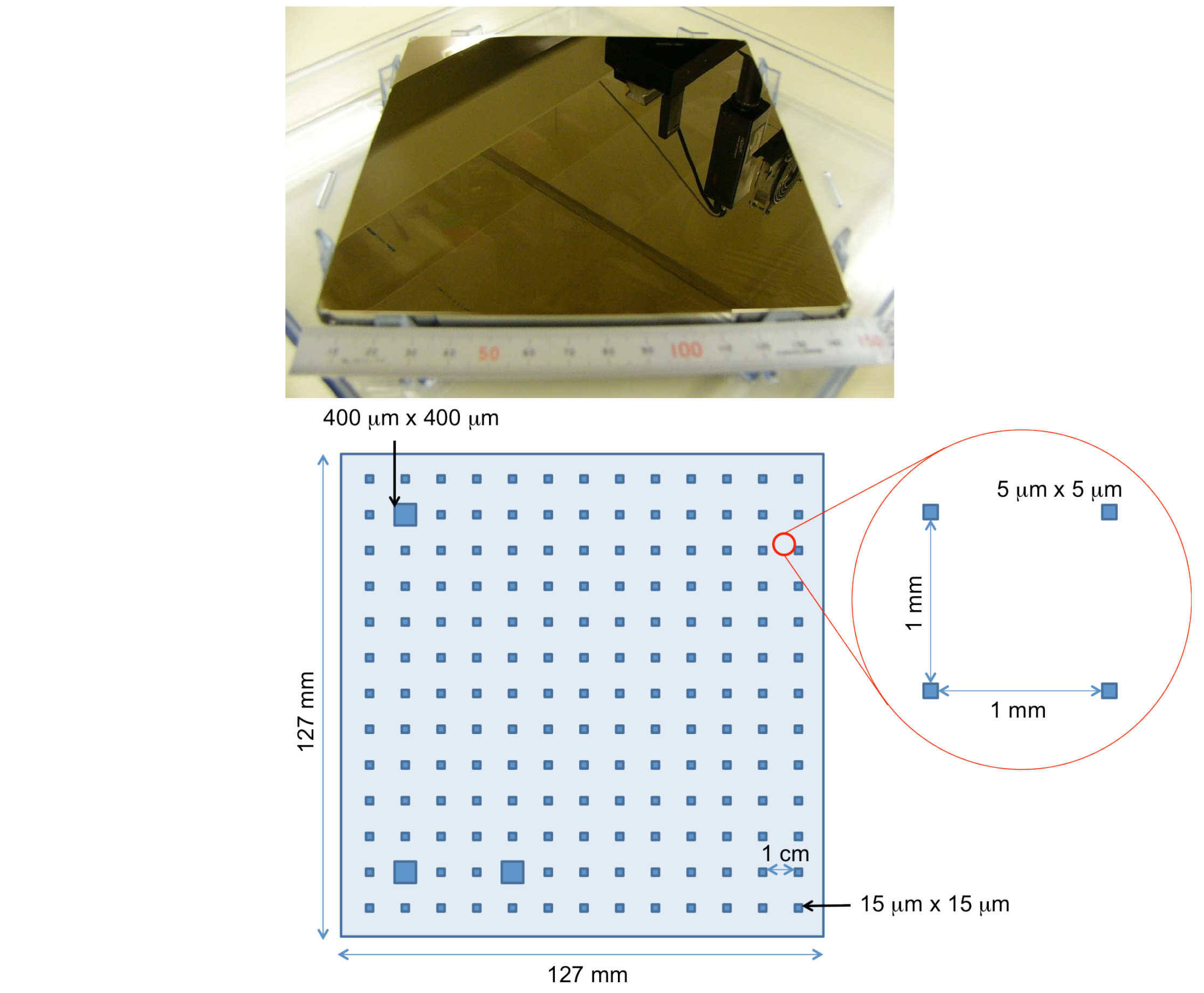}
        \caption{\emph {Photomask and its mark pattern~\cite{kimura}. Marks with different sizes are used: the 5 $\mu$m side square marks allow for the track position correction while the 15 $\mu$m and 400 $\mu$m marks are used as a positioning guide for the microscope.}}
        \label{photomask}
\end{figure}

	In order to match the geometry of the Moir\'e deflectometer, emulsion films with an area as large as 20 $\times$ 20 cm$^2$ will be employed. To maintain the track position accuracy at the micrometric level, the film expansion due to changes in the environmental conditions undergone by emulsions during the experiment has to be kept under control over that area. Furthermore, the expansion of the emulsion films is not uniform as the plastic base is affected by the gelatine layer shrinkage caused by the development process. For OPERA type emulsions, the typical size of the non-uniform expansion component, called "local expansion", is about 0.3 $\mu$m (10 $\mu$m) over a film area of several 100 $\mu$m$^2$ (100 cm$^2$). 
	To limit this effect the films can be produced by pouring the gel on a glass plate, since glass guarantees a better thermal and humidity stability with respect to plastic. Tests for the production of  emulsions on glass are presently being performed in Bern. Local expansions can be corrected for by printing a reference frame directly on the emulsion layer before the usual development process. This can be done with the contact printing method~\cite{kimura} allowing the imprinting on the films of reference marks from a light source with sub-micrometric accuracy. A photomask containing square holes (marks) with a side of 5 $\mu$m spaced by 1 mm is used to allow for the track position correction. Further square marks (with sides of 15 $\mu$m and 400 $\mu$m) facilitate the centering of the microscopic view on the 5 $\mu$m marks (figure~\ref{photomask}). 
	
	Since the reference marks are manufactured with a position accuracy of 0.1 $\mu$m, the printed marks act as reference points on the emulsion film. By comparing the measured coordinates with those defined by the marks on the film, one can correct for the local expansion.
	
\begin{figure}[h]
        \centering
          \includegraphics[height=10cm, width=13cm, angle=0,
          keepaspectratio]{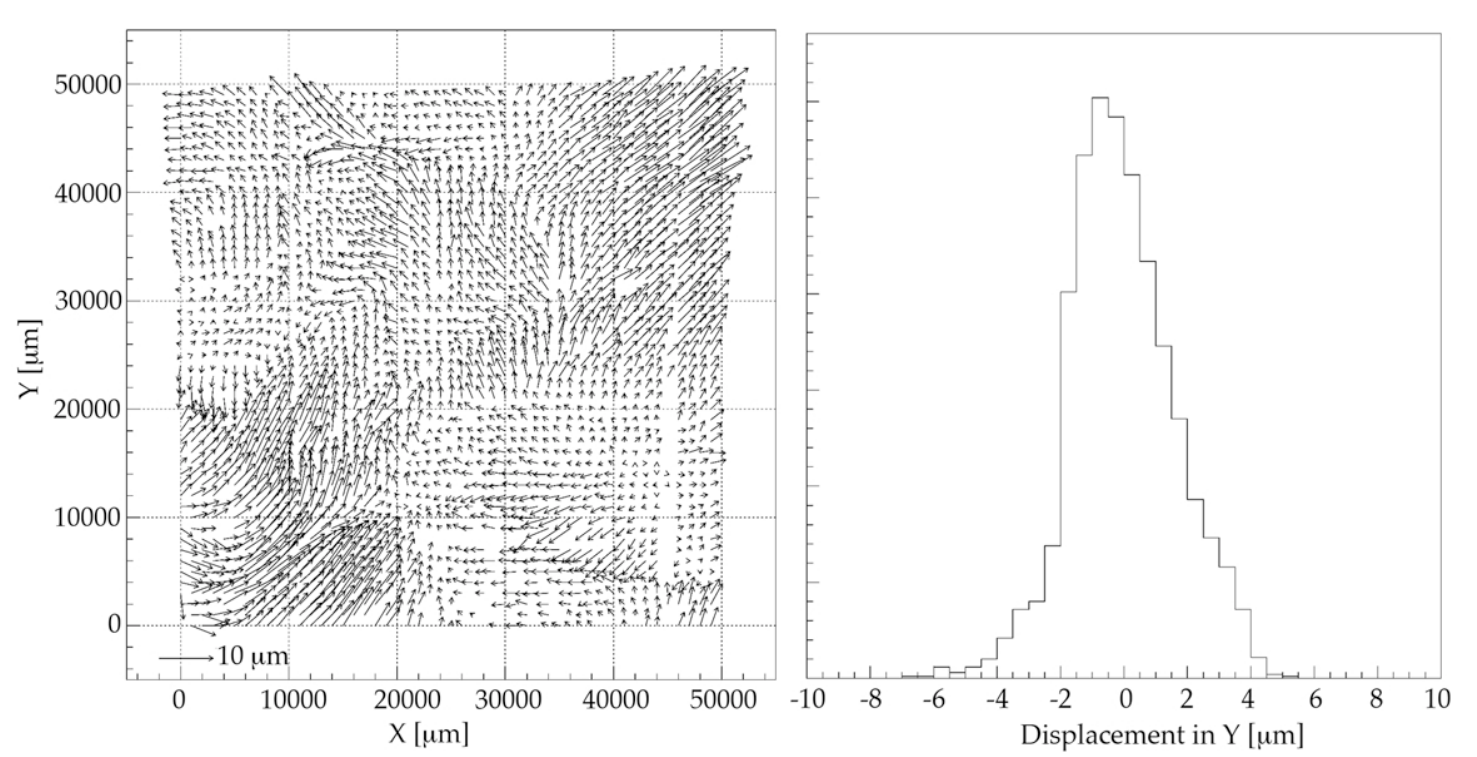}
        \caption{\emph {Local expansion evaluated for a 5 $\times$ 5 cm$^2$ OPERA film with the photomask pattern. The maximum displacement is 9.5 $\mu$m while the average is 2.8 $\mu$m (left). Y-projection distribution of the displacements (right).}}
        \label{localexp}
\end{figure}

	In the framework of the ESS an automatic recognition system of the reference marks was developed. The marked image is captured by a CCD camera. A reference square of 5 $\mu$m can be located with an accuracy of 0.3 $\mu$m.
	Figure~\ref{localexp} shows the results of a standard procedure for an OPERA film local expansion assessment. In the distortion map (left) the arrows indicate the displacements between the coordinates defined by the reference and measured marks. A maximum expansion of 9.5 $\mu$m is found, while the average displacement is 2.8 $\mu$m. In figure~\ref{localexp} (right) the Y-projection distribution of the displacements is also shown.

	An R\&D activity is also underway on the emulsion gel composition to improve the detector performance. We are presently producing and investigating new types of films with a gel developed at Nagoya University, also tuning the development process, and improving the sensitivity according to our specific experimental needs. For this purpose a high precision stage for pouring the custom-made gel on plastic (or glass) base supports was designed and constructed at LHEP. The level of the stage is tunable with a precision of the order of 0.1 mrad (a few 10 $\mu$m over 20 cm). It is equipped with a vacuum system and a soft silicone rubber packing to keep the flatness during film production.
First results on films produced in the underground LHEP laboratory are illustrated in Tab.\ref{tab1}, where the grain and fog density are reported for the new gels after different developing times and for a reference OPERA type film. 
	
\begin{figure}[htbp]
        \centering
          \includegraphics[height=7cm, width=14cm, angle=0,
          keepaspectratio]{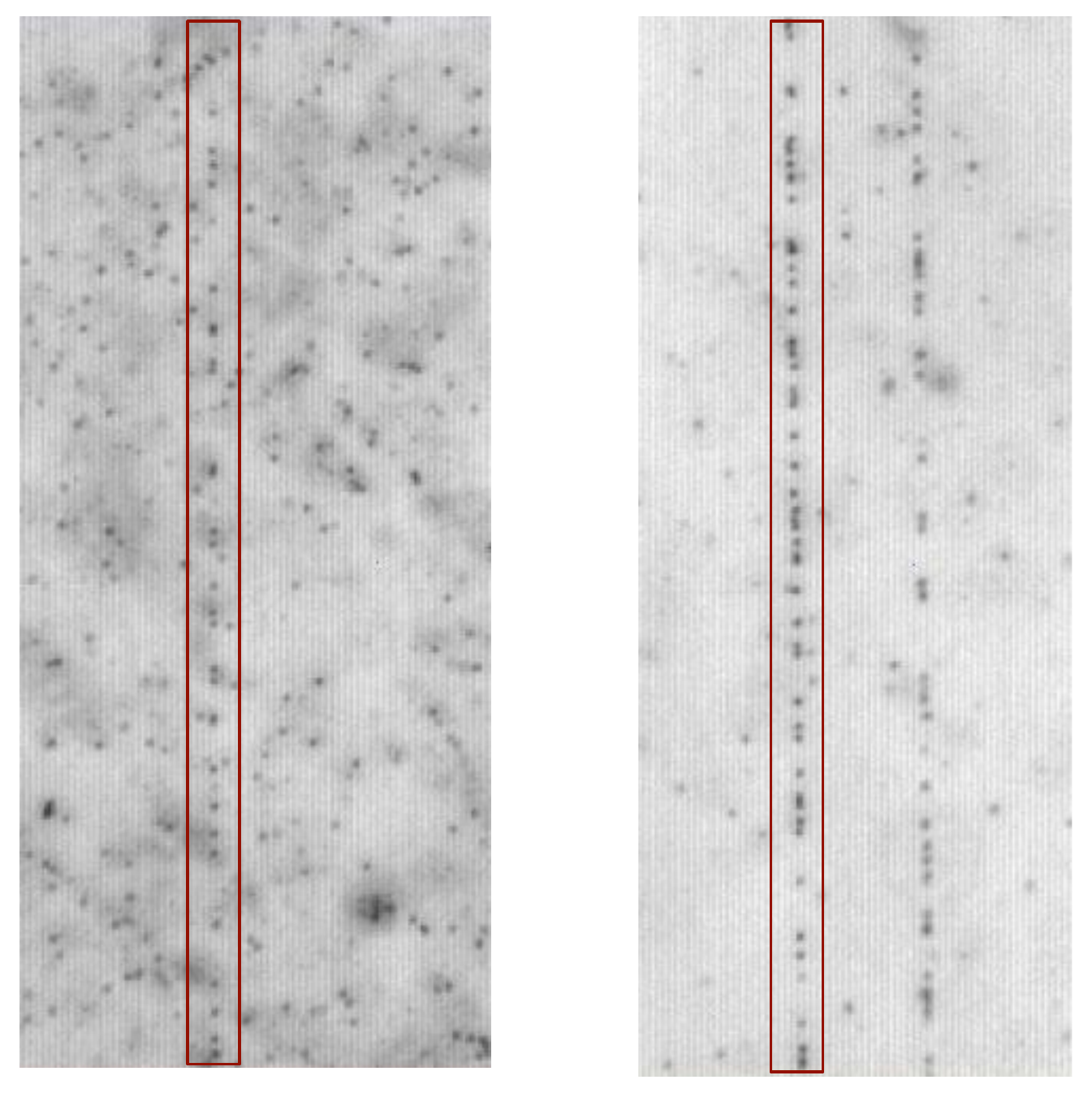}
        \caption{\emph{A MIP track in a reference OPERA film (left) and in a new gel film (right).}}
        \label{fignewgel}
\end{figure}

	In figure~\ref{fignewgel} a MIP track produced by a 10 GeV/c pion from the CERN PS/T9 beam-line is shown for a standard OPERA film and for the newly developed gel. The latter, realized by tuning chemical agents and characterized by a reduced gelatine content, indeed shows an increased sensitivity and a lower fog density. Further tests are being performed to establish the sensitivity of the new gel to MIPs.
\begin{table}[htbp]
	\caption{\emph {Sensitivity and fog density in the new gel film for different development times. The characteristics of the OPERA type films are also reported for comparison.}}
	\label{tab1}
\centering
\medskip
\begin{tabular}{cccc}\hline
\rule[.15in]{0.0in}{0.0in} Gel type & Development time & Grain density               & Fog density \\
                                                           &  [min]                         & [grains/100 $\mu$m] &  [grains/$\mu$m$^3$]  \\ \hline
OPERA     & 25 & 30.3 $\pm$ 1.6 &  $10.1~\pm$ 0.7 \\
Nagoya    & 20 & 47.7 $\pm$ 2.0 &   $~~1.9~\pm$ 0.2 \\
Nagoya    & 25 & 55.1 $\pm$ 2.6 &   $~~3.0~\pm$ 0.3 \\ \hline
\end{tabular}
\end{table}

\newpage

\section{Conclusions}
We propose to design and  operate a new position sensitive detector based on the emulsion film detector technique for the AEgIS experiment at CERN, aiming at the first measurement of the gravity acceleration of $\bar H$ atoms with a relative precision of 1\% on $\Delta g/ g$. With the emulsion detector we are proposing here one can reach a performance on the $\bar{H}$ annihilation point one order of magnitude better than what initially foreseen in the AEgIS proposal, thanks to the outstanding position resolution of emulsions, of the order of 1 $\mu$m.  The detector performance was evaluated by Monte-Carlo simulation showing that a 1 $\mu$m position resolution detector allows to achieve the same precision on $\Delta g/g$ as an ideal detector with infinite precision. With an optimized deflectometer with a pitch of 40 $\mu$m and 30\% open fraction, a precision of 0.5\% can be obtained with 1 $\mu$m position resolution, while the 1\% precision on $\Delta g/g$ is achieved with less than 2000 detected particles. This guarantees reaching the requested precision should the number of produced $\bar H$ atoms be smaller than expected. We are also studying other geometric configuration, e.g. a 20 $\mu$m pitch, to investigate possible further improvements.

	An intensive R\&D activity on emulsion films is in progress and will be further carried on since the proposed application implies the operation of emulsions in vacuum, for which no past experience exists. Tests are underway to study the emulsion operations in vacuum in terms of  mechanical stability, sensitivity and intrinsic background. Preliminary results show that the treatment of emulsion films with about 10\% glycerine solutions can prevent the formation of cracks due to vacuum conditions. The increased fog density observed in vacuum by employing reference OPERA film improves after the glycerine treatment from more than 60 grains per 1000 $\mu$m$^3$ to about 18 grains per 1000 $\mu$m$^3$. Furthermore, vacuum condition after glycerine treatment does not cause any appreciable deterioration of the crystal sensitivity, which was measured to be about 14\% for MIPs. New gels are also under study to produce emulsion films with improved sensitivity. The preliminary outcomes of the tests reported here are encouraging and set good prospects for the actual feasibility of our project.

\section{Acknowledgments}
	We wish to warmly acknowledge our technical collaborators: Jan Christen, Roger H\"anni, Roger Liechti, Pascal Lutz, Jacky Rochet, and Camilla Tognina.

\end{document}